\def\papertitle{Accurate Plate Reverb Parameter Estimation \\Using Two-Stage Evolutionary Search}
\def\paperauthorA{Byunghoo Park}
\def\paperauthorB{Jayeon Yi}
\def\paperauthorC{Takyoung Kim}
\def\paperauthorD{Minje Kim}

\documentclass[twoside,a4paper]{article}
\usepackage{etoolbox}

\usepackage[taskA]{dafx26challenge} 

\usepackage{amsmath,amssymb,amsfonts,amsthm}
\usepackage{siunitx}
\usepackage{euscript}
\usepackage[T1]{fontenc}
\usepackage[utf8]{inputenc}
\usepackage{ifpdf}
\usepackage[english]{babel}
\usepackage{caption}
\usepackage{subfig} 
\usepackage{color}
\usepackage{booktabs}
\usepackage{lipsum}

\usepackage{makecell}

\input glyphtounicode
\ninept

\newcounter{numauth}
\newcounter{listcnt}
\newcommand\authcnt[1]{\ifdefined#1 \stepcounter{numauth} \fi}

\newcommand\addauth[1]{
\ifdefined#1 
\stepcounter{listcnt}
\ifnum \value{listcnt}<\value{numauth}
\appto\authorslist{, #1}
\else
\appto\authorslist{~and~#1}
\fi
\fi}
\authcnt{\paperauthorB}
\authcnt{\paperauthorC}
\authcnt{\paperauthorD}
\authcnt{\paperauthorE}
\authcnt{\paperauthorF}
\authcnt{\paperauthorG}
\authcnt{\paperauthorH}
\authcnt{\paperauthorI}
\authcnt{\paperauthorJ}
\def\authorslist{\paperauthorA}
\addauth{\paperauthorB}
\addauth{\paperauthorC}
\addauth{\paperauthorD}
\addauth{\paperauthorE}
\addauth{\paperauthorF}
\addauth{\paperauthorG}
\addauth{\paperauthorH}
\addauth{\paperauthorI}
\addauth{\paperauthorJ}

\usepackage{times}

\newif\ifpdf
\ifx\pdfoutput\relax
\else
   \ifcase\pdfoutput
      \pdffalse
   \else
      \pdftrue
   \fi
\fi

\ifpdf 
  \usepackage[pdftex,
    pdftitle={\papertitle},
    pdfauthor={\authorslist},
    pdfsubject={Proceedings of the 29th International Conference on Digital Audio Effects (DAFx26)},
    colorlinks=false, 
    bookmarksnumbered, 
    pdfstartview=XYZ 
  ]{hyperref}
  \usepackage[pdftex]{graphicx}
\else 
  \usepackage[dvips]{epsfig,graphicx}
  \usepackage[dvips,
    pdftitle={\papertitle},
    pdfauthor={\authorslist},
    pdfsubject={Proceedings of the 29th International Conference on Digital Audio Effects (DAFx26)},
    colorlinks=false, 
    bookmarksnumbered, 
    pdfstartview=XYZ 
  ]{hyperref}
\fi
\usepackage[hypcap=true]{caption}
\title{\papertitle}

\affiliation
{\paperauthorA$^\ast$ \quad \paperauthorB$^\ast$ \quad \paperauthorC \quad \paperauthorD}
{\href{https://siebelschool.illinois.edu/}{Siebel School of Computing and Data Science} \\ University of Illinois Urbana-Champaign \\ Urbana, USA\\
{\tt \{bhpark3, jayeonyi, tk30, minje\}@illinois.edu}
}

\begin{document}
\ifpdf 
  \DeclareGraphicsExtensions{.png,.jpg,.pdf}
\else  
  \DeclareGraphicsExtensions{.eps}
\fi


\maketitle

\def\thefootnote{*}\footnotetext{These authors contributed equally to this work.}

\begin{abstract}
We describe our submission to Task A of the 1st DAFx parameter estimation challenge. The task is to recover the six physical parameters of a simulated metal-plate reverberator --- its dimensions and material properties --- from a single impulse response (IR). We treat this as a black-box optimization: candidate parameter sets are fed to the simulator and scored by a loss against the target IR. The method has two stages. The first uses CMA-ES, an evolutionary optimizer, to recover five of the six parameters, comparing IRs under an amplitude-normalized loss. Amplitude normalization makes the search robust but discards the cue to the sixth parameter, the plate's surface density; a second stage therefore estimates it alone, with a ternary search on the un-normalized loss. As the choice of loss strongly affects the search, we select it beforehand, and analyze why compression in the common multi-scale spectral loss degrades recovery. Finally, we test our method on a validation set of 50 IRs, discuss a pathological failure mode, and ablate to justify having two different stages instead of a unified CMA-ES search.
\end{abstract}

\section{Introduction}
\label{sec:intro}
A plate reverberator \cite{plate_reverb} produces artificial reverberation \cite{artificial_reverberation} by picking up the vibration of a driven metal plate; such units are now commonly emulated in software. Given a simulated plate IR, one may wish to recover the physical parameters that generated it, both to learn useful configurations and as a step towards broader creative applications.

The 1st DAFx parameter estimation challenge \cite{the_challenge} benchmarks this problem. It consists of two tasks: Task A, ``physical parameter identification'', is to infer the physical parameters of the plate from a simulated IR, while Task B, ``modal parameter estimation'', is to dissect the modes of vibration. In this work, we describe our submission to Task A.

Existing methods do not transfer readily. Xu and Ding \cite{xu2017parameter} adapt the Newton method to estimate coefficients of linear time-invariant systems, but it has been demonstrated only on simple third-order systems and does not readily scale to a problem of this size. Lee \emph{et al.} \cite{dar} optimize a differentiable room reverberator via gradient descent; our preliminary attempts to adapt this approach were unstable and prone to divergence, likely reflecting a rugged, ill-conditioned loss landscape. We therefore adopt the covariance matrix adaptation evolution strategy (CMA-ES) \cite{cmaes}, a derivative-free optimizer suited to exactly such problems. CMA-ES has also shown promise in similar parameter-estimation settings, such as articulatory speech synthesis \cite{pink_trombone_optimization}.

\section{Background}
\label{sec:background}

\subsection{Plate Reverb Simulator, and Task A of the Challenge}
\label{sec:plate}
The plate simulator \cite{the_challenge} implements the general solution of the damped Kirchhoff--Love equation \cite{plate_reverb, vibrationofplates}, governed by the set of physical parameters
\begin{equation}
    P := \{\rho, E, \nu, L_x, L_y, h, T_0, \eta_0, \eta_1, x_i, y_i, x_o, y_o\},
    \label{eq:params}
\end{equation}
which are, in order: the volumetric density ($\rho$), Young's modulus ($E$), Poisson's ratio ($\nu$), the width ($L_x$), the height ($L_y$), the thickness ($h$), the tension applied via connections to the chassis ($T_0$), decay constants ($\eta_0,\eta_1$), the input location $(x_i,y_i)$, and the output location $(x_o,y_o)$. We refer the reader to the challenge documentation \cite{the_challenge} for the full model. For our purposes, a relevant property is that the simulator output is a sum of second-order infinite impulse response (IIR) filters whose coefficients are functions of the physical parameters. We therefore estimate the parameters by optimizing $P$ directly, searching for the parameter set whose rendered IR best matches the target.

Task A of the challenge is to estimate $P_\text{true}:=\{x_o$, $y_o$, $L_y$, $\mu$, $T_0/\mu$, $D/\mu\}$, a derived subset of $P$ associated with a given IR, where $\mu:=\rho h$ is the surface density and $D:=Eh^3/12(1-\nu^2)$ is the flexural rigidity. Other parameters ($\nu, L_x, \eta_0, \eta_1, x_i, y_i$) remain fixed. This task design allows for a one-to-one correspondence between the IR and the six parameters, and thus a fair evaluation via normalized mean square error (NMSE):
\begin{equation}
    \text{NMSE}(P_\text{true},P_\text{est}) = \frac{1}{6} \sum_{i=1}^{6} \left( \frac{P_{\text{true}, i} - P_{\text{est}, i}}{M_i - m_i} \right)^2
    \label{eq:nmse}
\end{equation}
where $M_i$ and $m_i$ are the upper and lower bounds of the values each parameter may assume, while $P_\text{true}$ and $P_\text{est}$ are the 6-dimensional vectors of true and estimated parameters. The bounds are derived from the legal ranges of the raw quantities $E$, $\rho$, $h$, $L_y$, $T_0$, $x_o$, and $y_o$. 

\subsection{Covariance Matrix Adaptation Evolution Strategy}
\label{ssec:cmaes}
Covariance matrix adaptation evolution strategy (CMA-ES) \cite{cmaes} is a derivative-free optimizer for continuous black-box objectives. At each generation, it samples a population of candidate solutions from a multivariate Gaussian, evaluates them, and adapts the distribution: the mean is moved towards the rank-weighted best candidates, while the covariance matrix and a global step size are updated to align the search with the directions in which good candidates have consistently been found \cite{cmaes}.

CMA-ES is designed for ill-conditioned, non-separable problems with many local optima (i.e. ``multimodal'') \cite{cmaes}. Its covariance matrix learns the local geometry of the objective, providing implicit preconditioning of the search space, while its rank-based selection yields invariance under monotonic transformations of the objective. These properties suit the present problem: the inverse mapping from IR to parameters is also expected to be ill-conditioned and multimodal, the regime CMA-ES is built for. We note that, although the simulator is at least piecewise differentiable, direct gradient descent on it tended to diverge, and proved difficult to stabilize through hyperparameter tuning.

\section{Methodology}
\label{sec:methodology}
Our method estimates the six Task A parameters over two stages: a CMA-ES search over the seven raw physical parameters, followed by a dedicated refinement of the mass density $\mu$. 

To meet the runtime requirements set in place by the challenge \cite{the_challenge}, we truncated all IRs and IR generations to the first 0.25s, while retaining the sample rate of \SI{44.1}{\kilo\hertz}. Moreover, all IR syntheses during the CMA-ES searches were run on a Quadro RTX 6000 GPU using a PyTorch port of the plate. 

Both stages minimize a common objective: an $L_1$ distance on STFT magnitudes between the target IR and the one rendered by the plate simulator. The STFT uses a 4096-point Hann window with a hop of 1024. Frames are centered on the signal samples, with reflection padding applied at both boundaries. Contrary to the baseline solution \cite{the_challenge}, we do not log-compress the magnitudes nor use multiple resolutions, as these were either found to deter convergence or be of less help. See Sec.~\ref{sec:loss} for an analysis. 

\subsection{Stage 1: CMA-ES Search}
\label{ssec:stage1}

We found CMA-ES searches over a seven-dimensional raw-parameter space of $E$, $\rho$, $h$, $L_y$, $T_0$, $x_o$, and $y_o$ to recover all Task-A parameters well, with the exception of $\mu$. Therefore, in Stage 1, we launch multiple \emph{restarts} of said search, and upon the first success compute and fix the five parameters $x_o$, $y_o$, $L_y$, $T_0/\mu$, and $D/\mu$ using the outputted seven-parameter estimates.

A single \emph{restart} runs CMA-ES (Sec.~\ref{ssec:cmaes}), updating a population of seven-element vectors over successive generations until one of four conditions is met: a budget of $30{,}000$ loss function evaluations, a tolerance of $1\times10^{-5}$ on the range of recent best-loss values, or a per-generation or per-restart wall-clock limit ($300$ and $600$ seconds, respectively). Each candidate is rendered by the simulator and compared to the target. Both IRs are peak-normalized before a loss can be derived; this makes the loss comparable across different IRs and accelerates optimization (Sec.~\ref{sec:results}).

The restarts are launched sequentially until one succeeds. Every restart draws its initial distribution mean by Latin hypercube sampling (LHS) in a normalized $[-1,1]^7$ cube, to ensure diversity across restarts. CMA-ES then operates in this normalized space with initial step size $\sigma_0 = 0.6$, mapping affinely onto the physical bounds of $E,\rho,h,L_y,T_0,x_o,y_o$ to render a candidate. To reduce runtime across the entire search all restarts, Optuna~\cite{optuna} prunes unpromising restarts via a SuccessiveHalvingPruner (reduction factor $3$, min resource $15$) and selects each restart's population size in $[30,60]$, as we had observed large population sizes to be sometimes detrimental to convergence. As the ground truth is unavailable at inference time, the search is configured to halt once any completed restart reaches a loss below the empirically chosen threshold of $0.01$, or after $400$ unsuccessful restarts.


We chose not to pursue a direct search over the six-dimensional Task-A space, as preliminary experiments showed greatly reduced convergence. A good out-of-bounds handler would be necessary to leverage the complex bounds derived from the seven raw parameters (Sec.~\ref{sec:plate}). This question is left for future research.


\subsection{Stage 2: Surface-Density Refinement}
\label{ssec:stage2}
The peak normalization employed in Stage 1 removes the absolute amplitude, and with it the only cue to $\mu$. Thus, Stage 1 recovers $\mu$ poorly. Meanwhile, with the ratios $D/\mu$ and $T_0/\mu$ and the plate geometry held fixed, we found $\mu$ to be monotonic to the overall amplitude of the response. 

Therefore we estimate $\mu$ with a separate 50-iteration ternary search over $\mu$, holding the other five parameters at their Stage 1 estimates and minimizing the same objective without any IR normalization. The search spans $\mu \in [2.43, 106.15]$, the range induced by the bounds on $\rho$ and $h$.

\begin{table}[t]
\centering
\caption{Component decomposition from $L_1$-STFT to the MSS+SC variant (components defined in text); the challenge MSS baseline is the \emph{log + multi} row, and \emph{log (bounded)} is the variant $\log(1{+}x)$. Values are geometric-mean NMSE and convergence rate ($\text{NMSE}<0.02$) over the 50 IRs.}
\label{tab:loss_decomp}
\begin{tabular}{@{}l c S[table-format=1.2]@{}}
\toprule
Configuration & Geomean NMSE & {Conv.\ rate} \\
\midrule
$L_1$-STFT (base)      & \num{4.9e-14} & 1.00 \\
\quad + multi          & \num{4.8e-14} & 1.00 \\
\quad + SC             & \num{1.0e-13} & 0.98 \\
\quad + SC + multi     & \num{3.2e-14} & 1.00 \\
\midrule
\quad + log (bounded)  & \num{8.9e-4}  & 0.40 \\
\quad + log            & \num{1.3e-1}  & 0.00 \\
\quad + log + multi (MSS)   & \num{1.3e-1}  & 0.02 \\
\quad + log + SC       & \num{7.0e-3}  & 0.48 \\
MSS+SC (log + SC + multi) & \num{2.7e-4}  & 0.68 \\
\bottomrule
\end{tabular}
\end{table}

\begin{table}[t]
\centering
\caption{Effect of STFT window size on recovery, for the single-resolution uncompressed $L_1$-STFT loss. Recovery saturates by 4096 points (bold, our choice); every window still outperforms the compressed losses of Table~\ref{tab:loss_decomp}.}
\label{tab:loss_window}
\begin{tabular}{@{}c c S[table-format=1.2]@{}}
\toprule
Window (pts) & Geomean NMSE & {Conv.\ rate} \\
\midrule
512  & \num{5.9e-12} & 0.92 \\
1024 & \num{3.7e-12} & 0.88 \\
2048 & \num{1.8e-13} & 0.96 \\
\textbf{4096} & \textbf{\num{4.9e-14}} & {\bfseries 1.00} \\
8192 & \num{9.2e-14} & 0.96 \\
\bottomrule
\end{tabular}
\end{table}

\section{Loss Analysis}

\label{sec:loss}

The selected $L_1$-on-STFT-magnitude loss, under our setting, outperforms the challenge's multi-scale spectral loss (MSS) baseline~\cite{engel2020ddsp}, which is a log-magnitude $L_1$ loss summed over multiple STFT resolutions. In this section, we analyze this gap. 

Two components separate our loss from this baseline: log-magnitude compression (\emph{log}) and multi-scale averaging (\emph{multi}). Because spectral-convergence terms are common in such losses, we additionally test a spectral-convergence term (\emph{SC})~\cite{spectral_convergence} --- a normalized magnitude error that emphasizes high-energy regions --- as an augmentation; adding it yields the MSS+SC variant~\cite{yamamoto2020parallel}. We toggle each component in isolation, holding the STFT framing fixed.

The comparisons are run on a set of 50 generated IRs. Because the candidate losses span very different numerical ranges, the CMA-ES termination tolerance and early-stopping threshold were set relative to each loss's own range, rather than the fixed values used in the main pipeline (Sec.~\ref{ssec:stage1}), so that all candidates are compared under comparable settings; the selected $L_1$-STFT loss was re-run under this setting, so its figures here differ from those of Sec.~\ref{sec:results}. Table~\ref{tab:loss_decomp} reports the geometric-mean NMSE and the convergence rate (the fraction of IRs recovered to $\text{NMSE}<0.02$) for each configuration.

Two observations follow. First, \emph{log} compression is responsible for the gap. On its own it reduces convergence to none and raises the geometric-mean NMSE from roughly $10^{-14}$ to $10^{-1}$; the challenge MSS baseline, which adds \emph{multi} to \emph{log}, inherits this collapse almost entirely, recovering only $2\%$ of IRs. The \emph{SC} and \emph{multi} components are individually harmless, each leaving recovery at the level of the base loss; their only effect is to partly offset the damage from \emph{log}. Adding \emph{SC} to the baseline --- the MSS+SC variant --- raises convergence from $0.02$ to $0.68$ (geomean $2.7\times10^{-4}$), still far short of the uncompressed $L_1$-STFT.

Second, the effect is specific to the ``unbounded'' 
$\log(x)$ form. 
Schw\"ar and M\"uller~\cite{schwaer2023} identify a related deficiency in harmonic estimation and propose the bounded form $\log(1+x)$ together with a full configuration (Smooth MSS). We find both less damaging than plain log --- the bounded form reaches geomean $8.9\times10^{-4}$ and their full configuration geomean $7.9\times10^{-11}$ --- yet neither matches the uncompressed loss.

Window size only has a secondary effect. Table~\ref{tab:loss_window} varies the single-resolution window from 512 to 8192 points: recovery is weakest at the smallest windows and saturates by 4096, which motivates our choice. The effect is small, however, and bounded --- every window recovers far better than any compressed loss, with even the worst (512 points, geomean $5.9\times10^{-12}$) an order of magnitude better than Smooth MSS. This localizes the failure to log-magnitude compression rather than to spectral resolution.

\section{Method Validation and Shortcomings}
\begin{figure}[t]
    \centering
    \includegraphics[width=0.72\columnwidth]{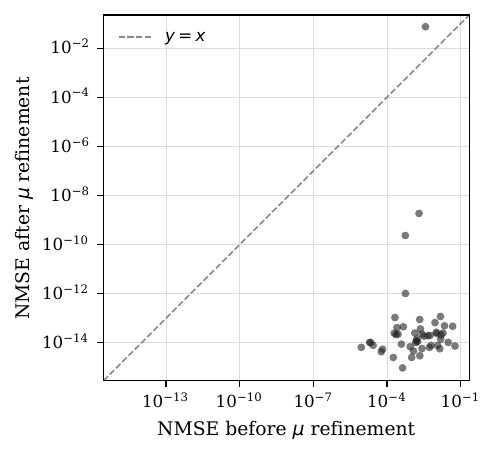}
    \caption{Per-IR NMSE (log--log) before and after Stage 2 $\mu$-refinement, over 50 generated IRs. All points except one (with pathological $x_o$) fall well below $y=x$, with refined NMSE often on the order of $10^{-14}$ or lower. }
    \label{fig:mu_refinement}
\end{figure}
\begin{figure}[t]
    \centering
    \includegraphics[width=0.9\columnwidth]{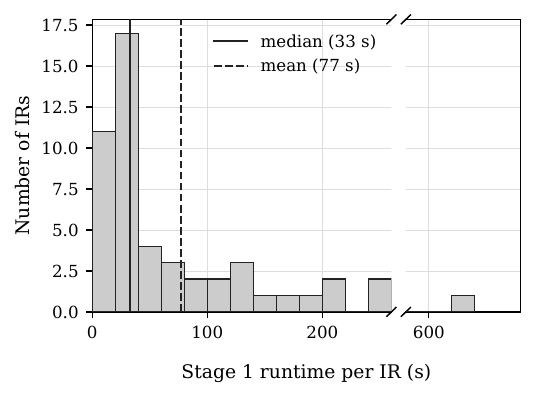}
    \caption{Distribution of per-IR Stage 1 runtime over the 50 validation IRs (linear axis, broken to show the tail). Most IRs finish near the median (32.6 s), while a few hard IRs extend past 200 s and a lone outlier reaches ${\sim}640$ s, pulling the mean (76.9 s) above the median.}
    \label{fig:runtime_hist}
\end{figure}
\label{sec:results}
To test the final two-stage design, we ran the full pipeline on another set of 50 IRs generated from the plate simulator with known, randomly sampled parameters. The two stages used no more than 4GB of GPU VRAM. 

Fig.~\ref{fig:mu_refinement} reports the NMSE of each IR before and after the Stage 2 ternary search. 49 out of 50 IRs improve: the median NMSE falls by eleven orders of magnitude, from $1.91\times10^{-3}$ after Stage 1 alone to $1.34\times10^{-14}$ after refinement. Likewise, the $L_1$-STFT loss on the unnormalized IRs falls by four or five orders of magnitude. Stage 1 takes a median of 32.6\,s and a mean of 76.9\,s per IR, after which Stage 2 contributes the NMSE improvement above at a negligible 0.62\,s per IR on average. Because Stage 2 adjusts only $\mu$, this also confirms that Stage 1 had mostly already recovered the other five parameters ($D/\mu$, $T_0/\mu$, $L_y$, $x_o$, $y_o$) to comparable precision.

The outlier in Fig.~\ref{fig:mu_refinement}, with its ground-truth $x_o=0.999980$ hovering very close to the upper limit of 1.0, represents a pathological corner case. While Stage 1 had estimated $x_o$ up to an impressive $\sim$0.02\% relative error ($x_o=0.999760$), the value of $(1-x_o)$ for the ground-truth and the prediction differed by approximately one order. Meanwhile, the simulator is known to set the amplitude of each mode proportional to $|\sin (x_o \pi m)|$, where $m$ is an integer associated with the x-axis component of each mode, typically $m \leq 200$. However, $|\sin (x_o \pi m)|$ is almost proportional to $(1-x_o)m$ when $x_om$ is close to any integer. Thus, for this particular IR, all mode amplitudes were altogether simultaneously ``mispredicted'' by approximately one order. This in turn caused the estimations for $\mu$, also an amplitude factor, to drift dramatically. Conversely, when $1-x_o$ and $1-y_o$ are not too small, $|\sin (x_o \pi m)|$ and $|\sin (y_o \pi m)|$ are thrown off by different magnitudes, and $\mu$ is correctly estimated. Meanwhile, the median estimation error for $x_o$ and $y_o$ were at the order of $10^{-8}$, with the $2 \times 10^{-4}$ error of the outlier being the top-1.

\begin{table}[t]
\setlength{\tabcolsep}{3pt}
\centering
\caption{Our two-stage pipeline vs.\ One-stage joint CMA-ES on the 50 validation IRs. Our pipeline is faster and more accurate. Both fail on the pathological outlier of Fig.~\ref{fig:mu_refinement}, with comparable error, but the one-stage search uses far more time.}
\label{tab:ablation}
\begin{tabular}{@{}lccc@{}}
\toprule
& \multicolumn{2}{c}{All 50 IRs}
& \makecell{Outlier in\\Fig.~\ref{fig:mu_refinement}} \\
\cmidrule(lr){2-3}
\cmidrule(l){4-4}
Search
&  \makecell{NMSE \\ median} 
& \makecell{Time (s)\\median / mean}
& \makecell{NMSE / Time (s)} \\
\midrule
Proposed
& \textbf{\num{1.3e-14}}
& \textbf{33.1 / 76.9}
& \num{7.6e-2} / 205 \\
One-stage
& \num{2.0e-14}
& 36.7 / 167
& \num{1.4e-2} / 3604 \\
\bottomrule
\end{tabular}
\end{table}

Observing the above, one may wonder if an alternate one-stage CMA-ES on unnormalized IRs would yield better estimations by jointly predicting $\mu$ and eliminating error propagation between stages (with the stages themselves). 
However, after scaling all termination criteria by each IR's peak amplitude for a fair comparison, we discovered this one-stage search to slightly fall behind in NMSE across the set, while being slower (Table~\ref{tab:ablation}). Although the pathological outlier is better estimated by the one-stage search, this estimation is produced only after hitting the restart cap at over an hour, versus a few minutes taken for the two-stage search. We therefore adopt the two-stage pipeline for our final submission.


\section{Conclusion}
\label{sec:conclusion}
We have described our submission to Task A of the 1st DAFx parameter estimation challenge, in which an IR from a plate simulator is inverted back to its six generating physical parameters. We adopted a two-stage evolutionary approach: a CMA-ES search that recovers five of the six target parameters ($D/\mu$, $T_0/\mu$, $L_y$, $x_o$, $y_o$) under an amplitude-normalized loss, followed by a ternary search that refines the surface density $\mu$ on the un-normalized loss. We found the choice of loss to be decisive: an uncompressed $L_1$-on-STFT-magnitude loss recovers the parameters almost perfectly, whereas the log-magnitude compression in the common multi-scale spectral loss collapses recovery. On a validation set of 50 IRs, our method reaches NMSE values of $\sim10^{-14}$ in about 80 seconds per IR, 
and an ablation confirms that the two-stage split enables faster and more accurate estimations compared to a single-stage joint CMA-ES search. We also characterized a pathological failure mode that arises when an output pickup coordinate lies extremely close to the plate boundary. Our solution has been run on the challenge's evaluation set; the results will be determined by the organizers.


\bibliographystyle{IEEEtranDAFx}
\bibliography{DAFx26_tmpl} 

@article{dar,
  title={Differentiable artificial reverberation},
  author={Lee, Sungho and Choi, Hyeong-Seok and Lee, Kyogu},
  journal={IEEE/ACM Transactions on Audio, Speech, and Language Processing},
  volume={30},
  pages={2541--2556},
  year={2022},
  publisher={IEEE}
}

@article{spectral_convergence,
  title={Fast spectrogram inversion using multi-head convolutional neural networks},
  author={Ar{\i}k, Sercan {\"O} and Jun, Heewoo and Diamos, Gregory},
  journal={IEEE Signal Processing Letters},
  volume={26},
  number={1},
  pages={94--98},
  year={2018},
  publisher={IEEE}
}

@techreport{the_challenge,
  title  = {{The 1st DAFx Parameter Estimation Challenge: A Benchmark for Plate Reverb System Identification}},
  author = {Ducceschi, Michele and Gabrielli, Leonardo},
  institution = {University of Bologna and Universit\`a Politecnica delle Marche},
  year   = {2025},
  note   = {Available at \url{https://github.com/LOGUNIVPM/1st-DAFx-Challenge}}
}

@inproceedings{optuna,
  title={Optuna: A next-generation hyperparameter optimization framework},
  author={Akiba, Takuya and Sano, Shotaro and Yanase, Toshihiko and Ohta, Takeru and Koyama, Masanori},
  booktitle={Proceedings of the 25th ACM SIGKDD international conference on knowledge discovery \& data mining},
  pages={2623--2631},
  year={2019}
}

@article{xu2017parameter,
  title={Parameter estimation for control systems based on impulse responses},
  author={Xu, Ling and Ding, Feng},
  journal={International Journal of Control, Automation and Systems},
  volume={15},
  number={6},
  pages={2471--2479},
  year={2017},
  publisher={Springer}
}

@article{plate_reverb,
  title={On the quality of plate reverberation},
  author={Arcas, Kevin and Chaigne, Antoine},
  journal={Applied Acoustics},
  volume={71},
  number={2},
  pages={147--156},
  year={2010},
  publisher={Elsevier}
}

@article{artificial_reverberation,
  title={Fifty years of artificial reverberation},
  author={Valimaki, Vesa and Parker, Julian D and Savioja, Lauri and Smith, Julius O and Abel, Jonathan S},
  journal={IEEE Transactions on Audio, Speech, and Language Processing},
  volume={20},
  number={5},
  pages={1421--1448},
  year={2012},
  publisher={IEEE}
}

@book{vibrationofplates,
  title={Vibration of plates},
  author={Leissa, Arthur W},
  volume={160},
  year={1969},
  publisher={Scientific and Technical Information Division, National Aeronautics and Space Administration}
}

@inproceedings{pink_trombone_optimization,
  title     = {{Optimization Techniques for a Physical Model of Human Vocalisation}},
  author    = {C\'amara, Mateo and Xu, Zhiyuan and Zong, Yisu and Blanco, Jos\'e Luis and Reiss, Joshua D.},
  booktitle = {Proceedings of the 26th International Conference on Digital Audio Effects (DAFx23)},
  address   = {Copenhagen, Denmark},
  pages     = {29--36},
  year      = {2023}
}

@article{cmaes,
  title={The CMA evolution strategy: A tutorial},
  author={Hansen, Nikolaus},
  journal={arXiv preprint arXiv:1604.00772},
  year={2016}
}

@inproceedings{engel2020ddsp,
  title={{DDSP}: Differentiable Digital Signal Processing},
  author={Engel, Jesse and Hantrakul, Lamtharn and Gu, Chenjie and Roberts, Adam},
  booktitle={International Conference on Learning Representations (ICLR)},
  year={2020}
}

@article{schwaer2023,
  author  = {Schw\"ar, Simon J. and M\"uller, Meinard},
  title   = {Multi-Scale Spectral Loss Revisited},
  journal = {IEEE Signal Processing Letters},
  volume  = {30},
  pages   = {1712--1716},
  year    = {2023},
  doi     = {10.1109/LSP.2023.3333205}
}

@INPROCEEDINGS{yamamoto2020parallel,
  author={Yamamoto, Ryuichi and Song, Eunwoo and Kim, Jae-Min},
  booktitle={ICASSP 2020 - 2020 IEEE International Conference on Acoustics, Speech and Signal Processing (ICASSP)}, 
  title={Parallel Wavegan: A Fast Waveform Generation Model Based on Generative Adversarial Networks with Multi-Resolution Spectrogram}, 
  year={2020},
  volume={},
  number={},
  pages={6199-6203},
  doi={10.1109/ICASSP40776.2020.9053795}}



\end{document}